\documentclass[aps,nolongbibliography, superscriptaddress, twocolumn]{revtex4-2}
\usepackage[british]{babel}
\usepackage{latexsym}

\usepackage[pdftex, colorlinks=true, linkcolor=red, citecolor=blue, urlcolor=magenta, pdfstartview=FitH]{hyperref}
\usepackage{graphicx}	
\usepackage{amssymb}
\usepackage{amsmath}
\usepackage{xcolor}
\usepackage[normalem]{ulem}

\pdfoutput=1

\usepackage{bm}
\usepackage{verbatim}
\usepackage{amsmath}
\usepackage{amssymb}
\usepackage[utf8]{inputenc}

\usepackage{nicefrac}
\usepackage{braket}
\usepackage{pdfcomment}
\usepackage{pbox}
\usepackage{makecell}
\usepackage{hhline}
\usepackage{multirow}
\usepackage{tabularx}
\usepackage{soul}
\usepackage{float}
\usepackage{microtype}
\usepackage[dvipsnames]{xcolor}

\begin{document}

\title{Emergence of Propagating Exciton-Polaritons\\ in Hybrid Waveguide-van der Waals Heterostructures}
%
%
\author{Alina Schubert}
\affiliation{Institute of Physics, University of Rostock, Rostock, Germany}
\author{Karoline Becker}
\affiliation{Institute of Physics, University of Rostock, Rostock, Germany}
\author{Andreas Thies}
\affiliation{Ferdinand-Braun-Institut, Leibniz-Institut für Höchstfrequenztechnik (FBH), Berlin, Germany}
\author{Rico Schwartz}
\affiliation{Institute of Physics, University of Rostock, Rostock, Germany}
\author{Takashi Taniguchi}
\affiliation{Research Center for Materials Nanoarchitectonics, NIMS, 1-1 Namiki, Tsukuba 305-0044, Japan}
\author{Kenji Watanabe}
\affiliation{Research Center for Electronic and Optical Materials, NIMS, 1-1 Namiki, Tsukuba 305-0044, Japan}
\author{Henije Stolz}
\affiliation{Institute of Physics, University of Rostock, Rostock, Germany}
\author{Alexander Szameit}
\affiliation{Institute of Physics, University of Rostock, Rostock, Germany}
\author{Matthias Heinrich}
\affiliation{Institute of Physics, University of Rostock, Rostock, Germany}

\author{Tobias Korn}
\email{tobias.korn@uni-rostock.de}
\affiliation{Institute of Physics, University of Rostock, Rostock, Germany}

\begin{abstract}
Integrating few-layer materials into photonic circuits is a promising concept for novel on-chip photonic applications. 
We incorporate transition metal dichalcogenides (TMDs) with femtosecond-laser-written surface waveguides, which are embedded in fused silica chips.
Our novel low-temperature optical spectroscopy setup enables coupling to the waveguide and simultaneous focus from the top to the TMD layer for a distinct excitation and collection of micro-photoluminescence (µPL) signals in several measurement geometries. 
Along these lines, we observe spectral changes of the A\,exciton for encapsulated TMD monolayers when capturing the µPL signal propagating through the waveguide. Depending on the thickness of the encapsulation with hexagonal boron nitride (hBN), these changes manifest as energetic redshifts of the A\,exciton, or even a splitting of the A\,exciton into two components.
We attribute this behavior to strong coupling of the waveguide mode and the exciton in the sample, giving rise to the formation of propagating exciton-polaritons. Our interpretation is supported by  calculations for a simplified model of a slab waveguide in the vicinity of an exciton by using the transfer matrix method.
Having proven to be a highly adaptable framework for the study of propagating polaritons, our experimental platform likewise holds great promise for harnessing the unique properties of exciton-polaristons in integrated photonic circuits. 
\end{abstract}

\maketitle
\section{Introduction}

Polaritons, the quanta of the hybridization of an electromagnetic and a polarization wave, represent the fascinating intersection between light and matter in the strong coupling regime \cite{klingshirn_semiconductor_2007}. As such, they are simultaneously the focus of intensive fundamental research and promising ingredients for future optoelectronic devices.
Fundamentally, the optoelectronic properties of semiconductors are defined by the formation of excitons, bound states of electrons in the conduction band and holes in the valence band \cite{klingshirn_semiconductor_2007}. In the strong coupling regime, so-called \textit{exciton-polaritons} emerge \cite{hopfield_theory_1958}. By virtue of this coupling, the dispersion relation for exciton-polaritons splits into anticrossing branches, the lower polariton (LP) and the upper polariton (UP) branch in the case of a single resonance. Notably, the introduction of photonic characteristics results in a striking reduction of effective masses and higher group velocities compared to uncoupled excitons, whereas the excitonic heritage of these hybrid entitites enables strong interactions and, by extension, exceptional nonlinear properties \cite{tassone_exciton-exciton_1999}. Hence, exciton-polaritons may be harnessed in applications ranging from efficient polariton lasers \cite{imamoglu_nonequilibrium_1996,kasprzak_boseeinstein_2006} to optical logic gates \cite{liew_optical_2008} and building blocks for quantum computation \cite{sanvitto_road_2016}.\\
As direct-bandgap semiconductors \cite{splendiani_emerging_2010,mak_atomically_2010} with exceptional exciton binding energies on the order of several 100\,meV\:\cite{chernikov_exciton_2014} and large oscillator strengths\:\cite{wang_colloquium_2018}, transition metal dichalcogenide (TMDs) monolayers constitute a particularly enticing class of host materials for the further investigation of exciton-polaritons. They have previously been harnessed to demonstrate the formation of cavity exciton-polaritons with micro-cavities \cite{liu_strong_2015,dufferwiel_excitonpolaritons_2015} and photonic crystals \cite{zhang_photonic-crystal_2018,gogna_photonic_2019}, as well as the formation of propagating exciton-polaritons in a slab waveguide \cite{kondratyev_probing_2023}.

Following the first experimental demonstration of exciton-polaritons in GaAs quantum wells within a Fabry-Pérot cavity, \cite{weisbuch_observation_1992}, modal confinement around an excitonic medium by optical resonators became the method of choice for accessing the strong coupling regime in the majority of studies \cite{kavokin_microcavities_2017}. Yet, early theoretical and experimental works also considered waveguides as an alternative means to suitably restrict the photon mode \cite{katsuyama_excitonic_1994}. In such systems, the coupling between the guided mode and quantum well excitons can drive the formation of propagating exciton-polaritons \cite{beggs_waveguide_2005, walker_exciton_2013}, which have also been demonstrated for etched rectangular and strip waveguides \cite{liran_fully_2018}.

\begin{figure*} [t!]
    \centering
    \includegraphics[width=1.0\linewidth]{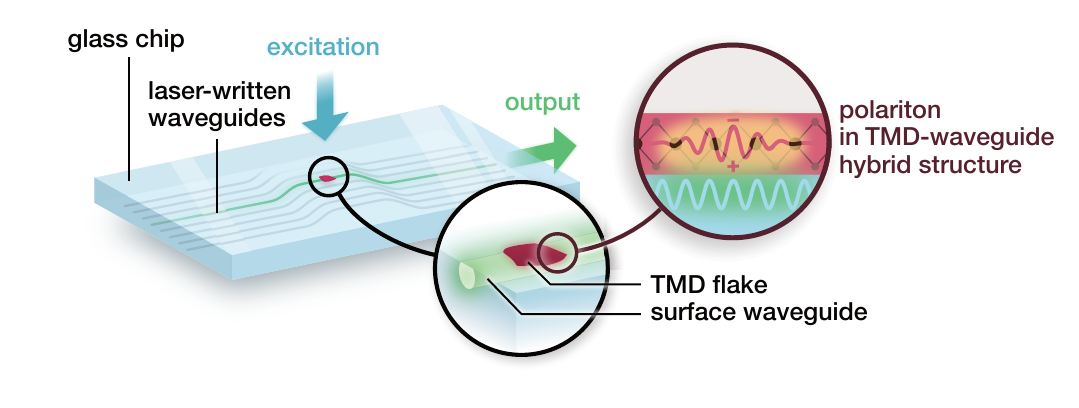}
    \caption{Schematic of the hybrid waveguide-TMD structure and the formation of propagating exciton-polaritons}
    \label{Panel1_introduction}
\end{figure*}

A particularly versatile technique for the realization of complex waveguide geometries is femtosecond laser direct writing (FLDW) \cite{davis_writing_1996, miura_photowritten_1997,szameit_discrete_2010}, where intense ultrashort laser pulses are focused into the volume of a transparent host material such as fused silica to create a permanent refractive index increase along arbitrary free-form trajectories within the bulk of the glass substrate. In turn, high-quality near-surface waveguide sections can be realized by subsequent polishing to the desired depth, enabling the localized and controlled overlap between the guided light and samples placed on top of the chip \cite{becker_polished_2025}. In particular, depositing a TMD monolayer on such a surface waveguide facilitates the interaction between excitons with the electromagnetic field of the waveguide mode and thereby the emission of light into the waveguide mode.

In this work, as sketched in Figure \ref{Panel1_introduction}, we investigate the optical properties of hybrid waveguide-TMD structures under cryogenic conditions. Depending on the monolayer's dielectric environment, which we control via encapsulation  with hexagonal boron nitride (hBN) of varying thickness, we observe a systematic redshift of the lowest-energy transition, the so called A\:exciton, and an eventual splitting into low- and high-energy components for emissions into the waveguide. We attribute this to the formation of propagating exciton-polaritons in the presence of strong coupling between the excitons in the TMD layer and the guided modes in the layered structure. Our versatile on-chip spectroscopy technique provides new perspectives on the control of propagating exciton-polaritons in laser-written surface waveguides. We demonstrate the potential of this hybrid architecture as a platform for exciton spectroscopy with readily-accessible in-plane geometries. At the same time, the feasibility of three-dimensional waveguide arrangements embedded within the volume of the chip paves the way towards innovative on-chip polaritonic optical devices with complex interconnects.

\section{Results and discussion}

\begin{figure*} [t!]
    \centering
    \includegraphics[width=1.0\linewidth]{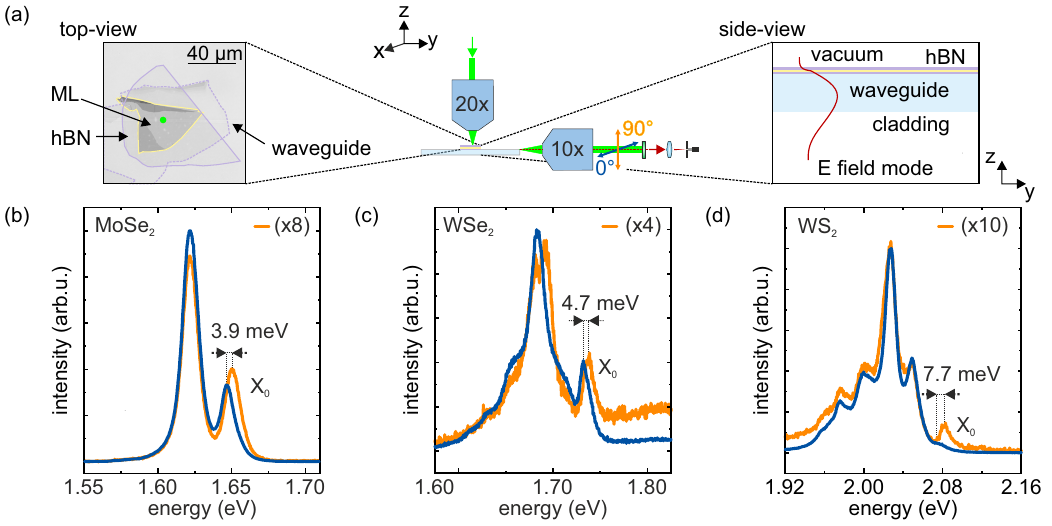}
    \caption{Low-temperature top excitation and side collection µPL measurements. (a) Schematic of the measurement geometry with adjustable detection polarization. In-plane polarization indicated blue, out-of-plane polarization indicated orange. 
    Further details on our setup, the surface waveguides and flake deposition can be found in Supplementary Figures~\ref{Supplement_fig_experimental_setup}, \ref{Supplement_fig_chip_principle} and  \ref{Supplement_fig_sample_pictures}, respectively. (b)-(d) Low-temperature µPL spectra for MoSe$_2$, WSe$_2$ and WS$_2$. The out-of-plane-polarized signal (orange) was scaled by the given factor for better comparison. The A\:exciton is marked by X$_0$.}
    \label{Panel2_TMD_spectra}
\end{figure*}

In a first set of measurements, we studied the low-temperature micro-photoluminescence (µPL) of three different TMD monolayers,  MoSe$_2$,  WSe$_2$ and WS$_2$, encapsulated in thin hBN and placed on a surface waveguide. As illustrated in Figure~\ref{Panel2_TMD_spectra}a, the flake is excited from above within its region of overlap with the surface waveguide. In turn, a Glan-Taylor prism serves to distinguish the in-plane and out-of-plane polarization components of the signal collected from this waveguide at the chip end facet. Figures~\ref{Panel2_TMD_spectra}b-d show the encapsulated-monolayer spectra of our three TMD materials. Note that the in-plane component is generally several times higher than the out-of-plane component, since the transition dipole of the bright A\:exciton is oriented in the $xy$ plane \cite{xiao_coupled_2012, zeng_valley_2012, mak_control_2012, schuller_orientation_2013} and the emission appears linearly polarized along $x$ when viewed from the side. The same applies to the so-called trion, a bound state of an exciton and an additional electron \cite{mak_tightly_2013}. In line with previous findings \cite{ross_electrical_2013}, we observe the exciton and trion peaks for MoSe$_2$ at 1.65\,eV and 1.62\,eV, respectively (Figure~\ref{Panel2_TMD_spectra}b). A direct comparison reveals that the spectral position of the trion is the same for both detection polarization directions. In contrast, the exciton peaks are shifted by 3.9\,meV to each other. This displacement is also noticeable for WSe$_2$ and WS$_2$, were the A\:exciton peak energy is located around 1.73\,eV and 2.08\,eV, despectively. (Figures~\ref{Panel2_TMD_spectra}c,d). Note that these tungsten-based TMDs exhibit a dark ground state \cite{zhang_experimental_2015, echeverry_splitting_2016}. Previous studies by Wang et al. and Wu et al. probed the dark, out-of-plane polarized exciton by focusing laterally onto the sample \cite{wang_-plane_2017,wu_up-_2020} and found its peak to be redshifted by approximtely 40\,meV to the bright A\:exciton. The high PL emission intensity reported in these studies could not be observed here, so that possible dark exciton contributions have comparable intensity to other low-energy emission features. The reason for this remains to be determined.

In contrast to the additional lower-energy peaks stemming from trions, biexcitons and localized defect states characteristic of tungsten-based TMDs \cite{huang_probing_2016,plechinger_identification_2015}, the spectrum of MoSe$_2$ only contains well-separated exciton and trion peaks, whose behavior we will further investigate in the following. Fundamentally, a number of factors can shift the energies of exciton peaks, e.g. temperature \cite{korn_low-temperature_2011}, changes in the dielectric environment \cite{florian_dielectric_2018}, strain \cite{castellanos-gomez_local_2013,conley_bandgap_2013} or doping \cite{mak_tightly_2013}. Absent these influences, the polarization-dependent difference of the A\:exciton energy and the conspicuous absence of similar shifts of the trion observed in our measurements must be of different origin. Along these lines, it is instructive to view the peak of the A\:exciton as superposition of two energetically close but differently polarized components: A lower-energetic contribution with higher intensity which is polarized in-plane, and a higher-energetic contribution with lower intensity that is either out-of-plane polarized or unpolarized background. This interpretation is supported by the observed angle dependence of the shift detailed in Supplementary Figure~\ref{Supplement_fig_MoSe2_angle_series}. The presence of such a splitting of the A\,exciton in our system therefore indicates the emergence of propagating exciton-polaritons between the encapsulated monolayer and the waveguide structure, as has previously been found in comparable systems \cite{liran_fully_2018, kondratyev_probing_2023}. This is consistent with the fact that MoSe$_2$ holds the lowest oscillator strength and shows the smallest shift of the three materials investigated here, whereas WS$_2$ holds the highest oscillator strength and shows the largest shift \cite{leppenen_exciton_2020}. The oscillator strength of single trions is much smaller than the single exciton oscillator strength \cite{combescot_trion_2003}. A more meaningful parameter, however, is the total oscillator strength per area, which is proportional to the absorption \cite{burstein_confined_1995}. The total trion oscillator strength scales with the charge carrier density \cite{glazov_optical_2020}. Hence, the total oscillator strength of the trion can be increased through doping, along with the associated light-matter coupling \cite{lundt_valley_2017}. MoSe$_2$ here exhibits a trion dissociation energy of 29.5\,meV measured in top-top geometry (cf. Supplementary Figure \ref{Supplement_fig_MoSe2_angle_series}b). Differential reflection measurements of comparable samples show no or only very weak absorption of the trion \cite{ross_electrical_2013,plechinger_trion_2016} This indicates that weak intrinsic doping and therefore a weak total trion oscillator strength are the reason why there is no apparent shift in the trion emission here.\\
Noteworthy to mention is that the exciton to trion intensity ratio appears to be influenced by the coupling to the waveguide, showing different behavior in top-side measurement geometry in comparison to top-top geometry. Hence, it should be avoided to consider the trion to exciton intensity ratio as a measure of doping \cite{cadiz_ultra-low_2016} in top-side geometry.

\begin{figure*} [t!]
    \centering
    \includegraphics[width=1.0\linewidth]{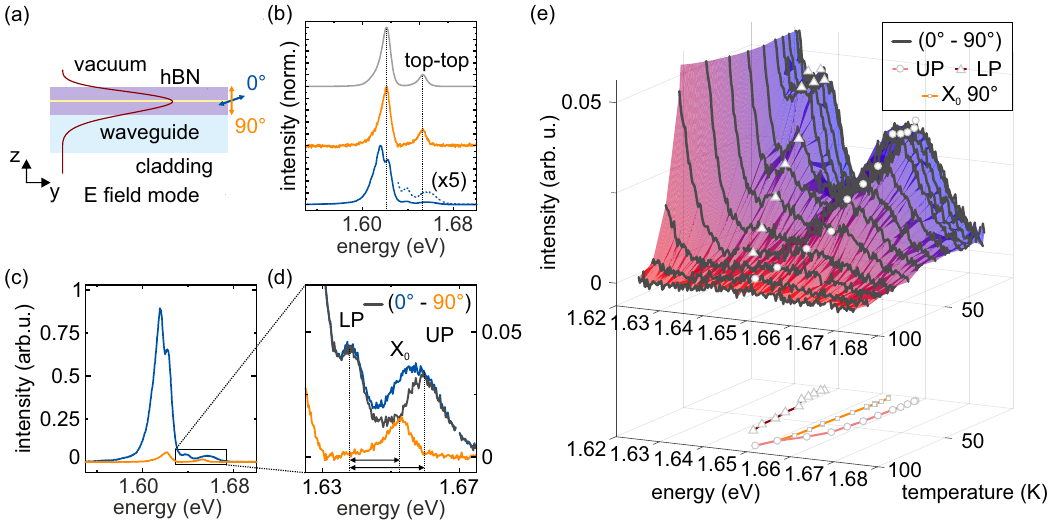}
    \caption{Low-temperature top excitation and side collection µPL measurements with MoSe$_2$ encapsulated in thicker hBN: (a) Schematic of the layered structure with thicker hBN and the mode of the electric field localized predominantly in the hBN rather than within the waveguide. (b) Spectra obtained with different detection polarization angles. In-plane polarization labeled blue with additional magnified A\:exciton, out-of-plane polarization labeled orange equivalent to the colors in Figure \ref{Panel2_TMD_spectra}. Comparison of the spectral form and the energy peak positions with top-top measurement geometry (grey).(c,d) Direct comparison of the spectra for in-plane and out-of-plane detection polarization with additional focus on the A\,exciton section. (e) Temperature series of in-plane spectra with subtracted out-of-plane signal focusing on the A\,exciton section with marked peak positions. Temperature-dependent energy shift of lower polariton, upper polariton and the uncoupled out-of-plane feature.}
    \label{Panel3_MoSe2_thick_hBN}
\end{figure*}

In order to investigate the emergence of the exciton-polaritons in our system in greater detail, we conducted PL measurements on MoSe$_2$ monolayers in thicker encapsulation. Figure~\ref{Panel3_MoSe2_thick_hBN} shows the results for an estimated hBN thickness of $\sim$40\,nm. This leads to wave guiding primarily in the hBN layers, as shown in Figure~\ref{Panel3_MoSe2_thick_hBN}a. We will discuss this further in the next section.\\
Compared to the previously investigated thin hBN encapsulation, these measurements yield in a surprisingly high trion-to-exciton intensity ratio, especially for horizontal detection polarization. In this measurement geometry, both trion and exciton split into two components, as shown in Figures~\ref{Panel3_MoSe2_thick_hBN}b-d, whereas detection in vertical polarization shows both exciton and trion lines as monolithic features. Since spectral form as well as positions of these peaks are equivalent to the ones obtained with top-excitation and top-collection measurement geometry, as shown in Figure~\ref{Panel3_MoSe2_thick_hBN}b, we interpret the latter as being associated with the emissions of uncoupled excitons from regions where the coherence length necessary for the formation of coherent exciton-polaritons falls below the wavelength, e.g. due to defects or impurities \cite{burstein_optical_1995}. Importantly, the selection rules of bright A\:excitons and trions in MoSe$_2$ \cite{xiao_coupled_2012, zeng_valley_2012, mak_control_2012, schuller_orientation_2013} preclude emissions with out-of-plane polarization, allowing us to leverage this signal to eliminate the unpolarized scattered-light background from our measurements, resulting in an even clearer view (black graph in Figure~\ref{Panel3_MoSe2_thick_hBN}d). The observed distinct splitting of $\sim$20.4\,meV in the thickly 
encapsulated MoSe$_2$ monolayer supports our hypothesis that exciton-polaritons indeed form within the integrated waveguide-TMD structures, causing the uncoupled exciton emission to bifurcate into an upper ($\sim$1.66\,eV) and a lower polariton ($\sim$1.64\,eV). We note that the spectra are comparable to the results by Dufferwiel et al. showing a splitting of the monolayer MoSe$_2$ A\:exciton in the same order of magnitude, albeit for cavity exciton-polaritons instead of propagating ones \cite{dufferwiel_excitonpolaritons_2015}.

In a further set of experiments, we turned our attention to the temperature dependence of the two polariton features. In line with established literature \cite{ross_electrical_2013}, the overall photoluminescence intensity increases dramatically towards lower temperatures (cf. Figure~\ref{Panel3_MoSe2_thick_hBN}e). Moreover, we find that the observed splitting likewise becomes more pronounced with decreasing temperature. To aid the distinction between exciton and polariton behavior, the peak position of the uncoupled exciton ($X_0~90^\circ{}$) is plotted here for comparison. As the bandgap narrows with increasing temperature, the energy of the uncoupled exciton decreases. In contrast, the energy of the lower polariton is clearly increasing, while the energy of the upper polariton decreases faster than the energy of the uncoupled exciton. In this vein, no  splitting was observable above 70\,K. Moreover, the peaks show the typical broadening at higher temperatures, and the concurrent reduction of the oscillator strength with temperature eventually leads to a transition into the weak coupling regime \cite{lundt_monolayered_2016}. WSe$_2$ encapsulated in thicker hBN showed comparable results. These are shown in the Supplementary Information in Figure~\ref{Supplement_Results_WSe2_thick_hBN}.

\begin{figure*} [t!]
    \centering
    \includegraphics[width=1.0\linewidth]{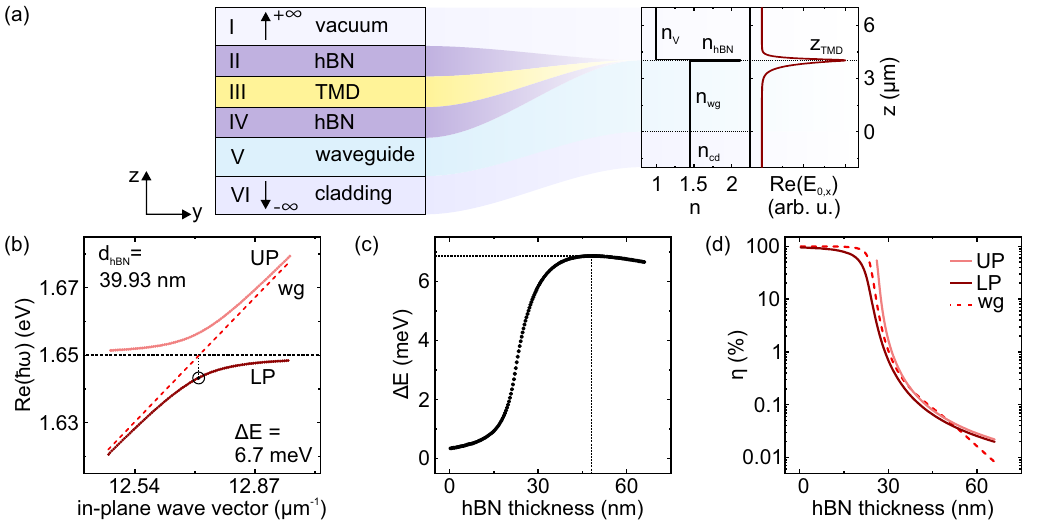}
    \caption{Slab waveguide model calculations of the integrated waveguide-encapsulated TMD-structure. (a) Left: Sketch of the idealized layered system with different thicknesses, refractive indices $n$ and excitonic properties in the TMD layer. Right: refractive index profile of the structure with electric field profile of the TE-mode of the lower polariton branch at the indicated bottleneck. (b)   
    Dispersion relation of the structure with 39.93\,nm hBN thickness of top and bottom hBN. bright red: stacked multilayer without TMD, dark red: lower polariton mode, bottleneck indicated with a circled black diamond, light red: upper polariton mode (c) Energy difference of the lower polariton branch at the bottleneck and the exciton resonance energy depending of the thickness of the hBN layers. (d) Logarithmically scaled overlap integral as a measure of coupling for the layered structure with varying hBN thickness and the eigenmode of the pure fused-silica waveguide.}
    \label{Panel3_Dispersion_Calculations}
\end{figure*}

Turning to the splitting behavior of the trion (Figure~\ref{Panel3_MoSe2_thick_hBN}b), we find that the higher-energetic feature measured in-plane coincides with the uncoupled features while the lower-energetic feature is not observed for uncoupled trions. This is further supported by the comparison of top-excitation-side-collection with the corresponding top-excitation-top-collection PL spectra (Supplementary Figure~\ref{Supplement_Results_MoSe2_scans_thick_hBN}). The scans reveal that the splittings of the exciton and of the trion indeed occur at the same excitation spots. They are most preferably measured for excitations in the vicinity of the waveguide and can be exclusively detected in-plane. Amounting to $\sim$5.3\,meV, the difference between the low-energy feature and the uncoupled trion 
is noticeably smaller than the respective displacement of the exciton. In addition, the trion splitting already vanishes between 20\,K and 30\,K, i.e. significantly earlier than that of the exciton (cf. Supplementary Figure~\ref{Supplement_Results_MoSe2_temperature_series}).\\

The spectrum measured in top-top geometry (Figure \ref{Panel3_MoSe2_thick_hBN}b) reveals a higher trion dissociation energy compared with the thin-encapsulated MoSe$_2$ sample. Due to stronger doping, the enhanced total trion oscillator strength may influence the formation of polaritons. Low temperatures additionally increase the oscillator strength of the trion \cite{arora_excitonic_2015}.
Trion-polaritons should not be considered independently but rather as a hybrid excitation of excitons, trions and photons \cite{dufferwiel_valley-addressable_2017,dhara_anomalous_2018,emmanuele_highly_2020,rana_exciton-trion_2021,koksal_structure_2021}. Unlike in the case of exciton-polaritons, as considered in the calculations given in the next paragraph, three branches arise as solutions for the dispersion relation of exciton-trion-polaritons, as long as the electron density is high enough. 
Due to the discrete mode of the cavity used in these previous works, trion and exciton resonances can be targeted directly, whereby these spectra are not directly comparable to the ones we show here.
Still, the observed peaks here could alternatively be addressed as lower polariton (1.616\,eV) below the trion, middle polariton (1.64\,eV) between trion and exciton and upper polariton (1.66\,eV).

For the purpose of simplicity, we consider only the exciton resonance to model the system theoretically. Based on the considerations of Beggs et al. \cite{beggs_waveguide_2005}, we describe the waveguide structure as a slab waveguide and use the transfer matrix method to calculate the dispersion relations of the layered structure including the excitons in the TMD layer. Details and parameters can be found in the methods section.\\
Since the slab waveguide model is highly simplified, our aim with this model is not to reproduce specific values, but rather to highlight trends that support our experimental findings and improve the basic understanding of the integrated waveguide-TMD-structure.\\
When assuming an hBN thickness of $\sim$40\,nm per layer, which is comparable to what we used in the experiment, the calculations result in a splitting of the lower polariton branch to the exciton resonance energy of 6.7\,meV, shown in Figure~\ref{Panel3_Dispersion_Calculations}b. That is in the approximate order of magnitude compared to the measured value of 13.9\,meV. Relaxation processes at the exciton-like section of the lower polariton branch create the highest population at the so-called bottleneck \cite{askary_exciton-polariton_1985}. Therefore the splitting is evaluated at that point. From the dispersion relation we calculate a group velocity of $v_\text{g}=0.28 \, c_0$ and a polariton lifetime of $\tau=50.3\,\text{ps}$, as shown in Figure~\ref{Supplement_fig_theory_group_velocity} in the Supplementary Information. This results in a propagation distance of $v_\text{g} \tau=4.26\,\text{mm}$. In the experiment a propagation distance of this magnitude would be limited by the size of the TMD sample.\\
Figure~\ref{Panel3_Dispersion_Calculations}a shows the stacking and the refractive index profile of the structure with the corresponding mode of the lower polariton branch at the bottleneck marked with a black diamond in Figure~\ref{Panel3_Dispersion_Calculations}b. It is apparent that the mode is guided in the hBN layer due to the very small difference in refractive index between waveguide and cladding together with the comparably high refractive index of hBN. The in-plane wave vector $k_y$ in the vicinity of the exciton resonance for $\sim$40\,nm hBN thickness does not meet the requirements for wave guiding inside of the actual waveguide ($n_\text{cd}k_0<k_y<n_\text{wg}k_0$ with the in-plane wave vector $k_y$, the wave vector in vacuum $k_0$ and the refractive indices $n$ for the waveguide (wg) and the cladding (cd) layers), because $k_y>n_\text{wg}k_0$. Therefore the mode decays exponentially in the waveguide. The energy difference $\Delta E=E_X-E_\text{LP}$ of the uncoupled exciton energy $E_X$ and the lower polariton energy $E_\text{LP}$ at the bottleneck increases strongly with increasing hBN layer thickness and achieves a maximum with an hBN thickness of 48.16\,nm for each layer, as shown in Figure~\ref{Panel3_Dispersion_Calculations}c. Subsequently, the splitting sightly decreases. This calculation supports our observation of a rather small splitting when the TMD sample is encapsulated in thin hBN and a larger splitting when encapsulating the sample with thicker hBN. Furthermore, this is an additional indication that the change of the exciton peak position in Figure~\ref{Panel2_TMD_spectra} is indeed because of the formation of polaritons.\\
Because the hBN thickness influences the mode significantly, there is a strong mismatch of the in-plane wave vectors and the modes of the layered structure with hBN and the pure surface waveguide. The eigenmode of the stacked structure needs to couple with the pure waveguide to provide a detectable signal at the end facet of the waveguide. The overlap integral, displayed in Figure~\ref{Panel3_Dispersion_Calculations}d, shows that from an hBN thickness of $\sim$20\,nm on the coupling efficiency decreases rapidly. For an hBN thickness of $\sim$40\,nm the coupling efficiency is reduced to 0.1\%. This can be also observed in our measurements. To collect the same quantity of counts, the sample with thin hBN layers required only a third of the excitation power at a ten times shorter integration time in comparison to the sample with thick hBN encapsulation.\\
The upper polariton branch cannot fully develop below a certain hBN thickness. For thin hBN thickness the exciton-like solutions of the upper polariton would hold $k_y<n_\text{cd}k_0$ which violates the condition for total internal reflection at the waveguide-cladding interface. Therefore, there are no exciton-like solutions with thin hBN, as shown for an hBN thickness of $\sim$8\,nm in Figure~\ref{Supplement_fig_theory_dispersion} in the Supplementary Information. This could be the reason why no upper polariton can be detected for the thin hBN sample.\\
Figure~\ref{Supplement_fig_theory_dispersion} also displays the dispersion relations for WSe$_2$ and WS$_2$ for an hBN thickness of $\sim$40\,nm. The hBN thickness and the radiative broadening influence the splitting the most. Other waveguide thicknesses, an additional cladding layer between waveguide and hBN or a different ratio of the non-radiative and the radiative broadening have only small influence, as shown in Figure~\ref{Supplement_fig_theory_thickness} in the Supplementary Information.

\section{Conclusion and outlook}
We demonstrated the integration of fused silica femtosecond laser direct written waveguides and hBN-encapsulated TMD monolayers. The integration opens up new opportunities for optical spectroscopy and extends the range of µPL measurement geometries, as excitation with and collection of light polarized perpendicular to the TMD plane is possible using the waveguide mode. We observe features of propagating exciton-polaritons in low-temperature µPL measurements, indicating strong coupling up to a temperature of 70\,K for MoSe$_2$. Supporting calculations of a slab waveguide model highlight the impact of the hBN thickness used for encapsulation on the splitting of the polariton branches.\\
TMDs and other nanomaterials offer vast possibilities of combinations and properties. Fused silica surface waveguides offer a flat and inert surface and thus serve as a durable platform. Accordingly, the integration of both in our hybrid TMD-waveguide structures paves the way for using propagating exciton-polariton effects in integrated photonics. Moreover, the hybrid platform more generally allows for investigating and harvesting the fascinating optical properties of any layered materials using waveguides.

\section{Methods}
\subsection{Sample fabrication}
    The surface waveguides utilized here were created by femtosecond laser direct writing with ultrashort pulses in a fused silica chip. The chip length is 5\,cm and the refractive index of the material is $n_0\,=\,1.45$. The waveguides show a $\Delta n \approx 10^{-3}$ higher refractive index and act as single mode waveguides. 
    The waveguides are written into the fused silica bulk material in a cascade pattern so that they have different levels of depth in the material. 
    Each waveguide is written deep into the fused silica bulk material  at the end facets, but is bent to define a region in the center of the chip that is $\sim$\,40\,µm above the baseline. A 2\,mm wide plateau serves as a potential surface waveguide section. 
    By careful polishing of the original surface, the surface sections of the waveguides are exposed. The surface waveguides utilized here are described in more detail by Becker et al. \cite{becker_polished_2025}.
    To ensure a better orientation throughout the chip, marker structures with a higher contrast were integrated into the chip to clearly assign the waveguide number, the surface region and the depth of the waveguide in the material to every waveguide.\\
    \newline
    Thin layers of 2D materials were separated from bulk crystals using mechanical exfoliation in ambient conditions and transferred to polydimethylsiloxane (PDMS) \cite{castellanos-gomez_deterministic_2014}. TMD monolayers were encapsulated in thin hBN by transferring from PDMS to PDMS. 
    The exfoliated TMD monolayers were obtained from bulk crystals by HQ Graphene.
    In the last step, the heterostructures were transferred to the surface waveguide by viscoelastic transfer \cite{castellanos-gomez_deterministic_2014}. To ensure sufficient visibility of the laser-defined marker structures for alignment of the 2D layers, the deterministic transfer setup utilized a quasi-transmission illumination geometry. For this, a mirror was placed beneath the sample holder.

\subsection{Optical spectroscopy}

Micro-photoluminescence (µPL) measurements were performed in several measurement geometries. Two long working distance microscope objectives were focused on front and end facet of the fused silica chip and one more on the TMD sample above the surface waveguide. This enabled µPL measurements with several combinations of top and side excitation or detection.
A custom-made cryostat hood with three windows for an AttoDry800 cryostat allowed measurements down to $\sim$ 4\,K.
The signal is collected with a fiber and analyzed with a Teledyne Princeton Instruments Acton spectrometer and a PIXIS charge-coupled device camera. 
A detailed representation of the experimental setup is shown in the Supplementary Information in Figure~\ref{Supplement_fig_experimental_setup}.
All measurements were conducted with a 532\,nm cw laser with varying powers and integration times depending on the measurement. 

\subsection{Transfer Matrix Method}

To describe the combination of the waveguide embedded in fused silica and covered with an encapsulated TMD layer, we used the transfer matrix method to calculate the waveguide modes \cite{born_principles_2019}. 
We consider transverse electric (TE) polarized waves and assume periodicity in time. The derivation of the transfer matrix method is based on Maxwell's equations and uses the fact that the tangential components of the electric and magnetic fields remain constant at the transition between two materials \cite{born_principles_2019}. 
The tangential components of the wave are grouped together in a vector:
\begin{equation}
    \begin{pmatrix}
    E_{0,x}(z_0)  \\
    H_{0,y}(z_0)   
    \end{pmatrix}=\mathbf{M_i}
    \begin{pmatrix}
    E_{0,x}(z)  \\
    H_{0,y}(z)   
    \end{pmatrix}
    \label{eq_tangential_progagation}
\end{equation}

The propagation of an electromagnetic wave through each layer $i$ is described by a transfer matrix $\mathbf{M_i}$ which is characteristic to the medium of each layer. 
For a homogeneous non-absorbing dielectric layer one has:
\begin{equation}
    \mathbf{M_i}= 
    \begin{pmatrix}
    \cos(k_z d)   &-\frac{i}{p}\sin(k_z d)   \\
    -ip\sin(k_z d)            &\cos(k_z d) 
    \end{pmatrix}
    \label{eq_M_homogenpus_film}
\end{equation}
The thickness of the layer is specified as $d$, $p$ is defined as $p=n \cos(\theta)$ with $n$ the refractive index of the layer and $\theta$ the angle of incidence, $k_z$ is the z-component of the wave vector $k_z=k \cos(\theta)$.

The transfer matrix method enables an easy description of the stacking by just multiplying another matrix for another layer. 

\begin{table}[t!]
\caption{Applied parameters for transfer matrix calculations \footnote{the refractive indices are labeled with $n$, the thicknesses with $d$. $E_A$ is the A\,exciton energy, $\tau_0^{LT}$ the radiative lifetime of the exciton at low temperatures and $\Gamma_{NR}$ the non-radiative broadening. Further comments can be found in section \ref{subsec_explanations_parameters} of the Supplementary Information.}}
\label{tab_calculation_parameters}
\begin{ruledtabular}
\begin{tabular}{lll}
$n_{cd}$                    &               & 1.45 \cite{becker_polished_2025}                  \\
$n_{wg}$                    &               & n$_{cd}$+10$^{-3}$ \cite{becker_polished_2025}    \\
$d_{wg}$                    &µm             &4                                                  \\
$n_{hBN}$                   &1              &2.1041 (MoSe$_2$) \cite{lee_refractive_2019}       \\
                            &               &2.1081 (WSe$_2$)                                   \\
                            &               &2.1288 (WS$_2$)                                    \\
$d_{hBN}$                   &nm             &8                                                  \\
$E_A$ MoSe$_2$              &eV             &1.65                                               \\
$\tau_0^{LT}$ MoSe$_2$      &ps             &5  \qquad \cite{Palummo_exciton_2015}              \\
$E_A$ WSe$_2$               &eV             &1.73                                                \\
$\tau_0^{LT}$ WSe$_2$       &ps             &3.8  \qquad \cite{Palummo_exciton_2015}            \\
$E_A$ WS$_2$                &eV             &2.08                                               \\
$\tau_0^{LT}$ WS$_2$       &ps             &2.3  \qquad \cite{Palummo_exciton_2015}            \\
$\Gamma_{0}$                &s$^{-1}$       &$(2\tau_0)^{-1}$ \qquad \cite{kavokin_cavity_2003} \\  
$\Gamma_{NR}$               &s$^{-1}$       &0.2$\Gamma_{0}$                                    

\end{tabular}
\end{ruledtabular}
\end{table}

The matrix describing the absorption by the exciton is given as follows \cite{beggs_interaction_2004}:
\begin{equation}
    \mathbf{M_X}= 
    \begin{pmatrix}
    1   &0   \\
    -2p\frac{r_{X}}{1+r_X}            &1
    \end{pmatrix}.
    \label{eq_M_exciton}
\end{equation}
The fraction containing the reflection coefficient $r_X$ for the TMD can be expressed as:
\begin{equation}
    \frac{r_{X}}{1+r_X}=i\frac{n}{p} \cdot \frac{\Gamma_0}{\tilde{\omega_0}-\omega-i\Gamma_{NR}}
\end{equation}
with the resonance frequency $\tilde{\omega_0}$, the frequency $\omega$, the exciton radiative broadening $\Gamma_0$ and the exciton non-radiative broadening $\Gamma_{NR}$.\\
We aim for the in-plane components of the wave vector $k_y$ and the eigenfrequencies of the system. Therefore, we only assume boundary conditions for an outgoing wave to the left $l$ and to the right $r$ \cite{beggs_waveguide_2005}.
\begin{equation}
    A
    \begin{pmatrix}
    1  \\
    -p_l 
    \end{pmatrix}=\mathbf{M}
    \begin{pmatrix}
    1  \\
    p_r 
    \end{pmatrix}
    \label{eq_M_exciton}
\end{equation}
$A$ is the relative amplitude of the fields and can be eliminated. The resulting eigenequation depends on the wave vector $k_y$ and the frequency. We set the values for the wave vector $k_y$ and numerically solve the equation 

\begin{equation}
    F(\omega)=M_{11}+M_{12}p_r+M_{21}\frac{1}{p_l}+M_{22}\frac{p_r}{p_l}=0
\end{equation}

to gain the eigenfrequencies. Like this the dispersion relations for the layered system can be obtained.\\
We consider a system as shown in Figure~\ref{Panel3_Dispersion_Calculations}a with the following layer order: Waveguide cladding, waveguide, hBN, TMD, hBN and vacuum. 
The parameters used for the calculation can be found in Table \ref{tab_calculation_parameters}. Further information regarding the determination of these parameters can be found in the Supplementary Information in section \ref{subsec_explanations_parameters}.\\
The coupling efficiency was calculated with \cite{saleh_fundamentals_2007, snyder_optical_2000, marcuse_theory_1991}:
\begin{equation}
    \eta=\frac{k_{y,pure}}{k_{y,layered}}\frac{|\int  E_{layered} E_{pure}^* dz|^2}{\int |E_{pure}|^2 dz \int |E_{layered}|^2 dz}
\end{equation}

\section{Acknowledgements}
The authors gratefully acknowledge J. Kuhlke for the assistance assembling the setup, as well as S. Scheel for fruitful discussions regarding the calculations.\\
The authors acknowledge financial support by the DFG \emph{via} SFB1477 (project No. 441234705).\\
T.K. acknowledges financial support by the DFG \emph{via} SPP2244 (project No. 443361515).\\ 
K.W. and T.T. acknowledge support from the CREST (JPMJCR24A5), JST and World Premier International Research Center Initiative (WPI), MEXT, Japan.\\

\bibliographystyle{apsrev4-2}
\bibliography{01PolaritonPaper}

\maketitle
\onecolumngrid
\newpage
\renewcommand{\thefigure}{S\arabic{figure}}
\renewcommand{\thetable}{S\arabic{table}}
\setcounter{figure}{0}
\setcounter{table}{0}
\section{Supplementary experimental data}

\begin{figure*} [h]
	\centering
	\includegraphics[width=1.0\textwidth]{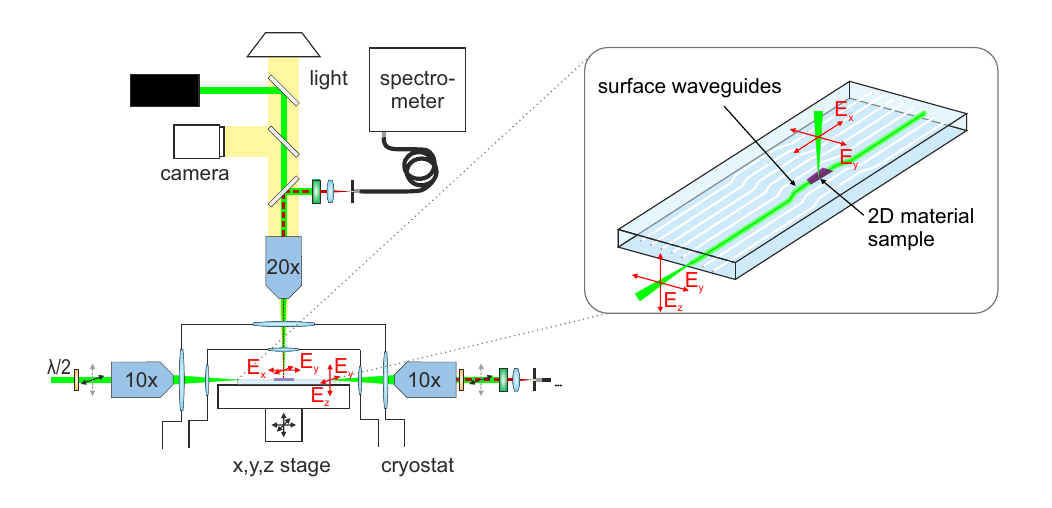}
	\caption{Schematic representation of the experimental setup. Display of the most relevant components of the micro-photoluminescence setup and the light path through the surface waveguide. The polarization direction planes of the PL signal are indicated by red arrows. The polarization direction planes that can be adjusted by the polarization optics of the setup are defined by black and grey arrows. The inset shows the cascade-like waveguide structure in the chip and the placement of the sample.}
	\label{Supplement_fig_experimental_setup}
\end{figure*}

\noindent
Measurement geometries, as shown in Figure~\ref{Supplement_fig_experimental_setup}:
\begin{itemize}
    \item top excitation - top collection
    \begin{itemize}
        \item PL measurements in regular micro-PL geometry
    \end{itemize}
    \item side excitation - top collection
    \begin{itemize}
        \item coupling into the surface waveguide and excitation of the 2D material via the evanescent field
        \item collection of the excited luminescence with the top objective focused on the overlap of waveguide and sample
        \item control over the excitation of the sample via a half-wave plate
    \end{itemize}
    \item side excitation - side collection
    \begin{itemize}
        \item coupling into the surface waveguide and excitation of the 2D material via the evanescent field
        \item collection of the luminescence from the end facet
        \item control over the detection polarization with a polarizing prism
    \end{itemize}
    \item top excitation - side collection
    \begin{itemize}
        \item excitation of the sample overlapping with the waveguide
        \item collection of the luminescence from the end facet
        \item control over the detection polarization with a polarizing prism
    \end{itemize}

\end{itemize}

\newpage
\begin{figure*}[h]
	\centering
	\includegraphics[width=1.0\textwidth]{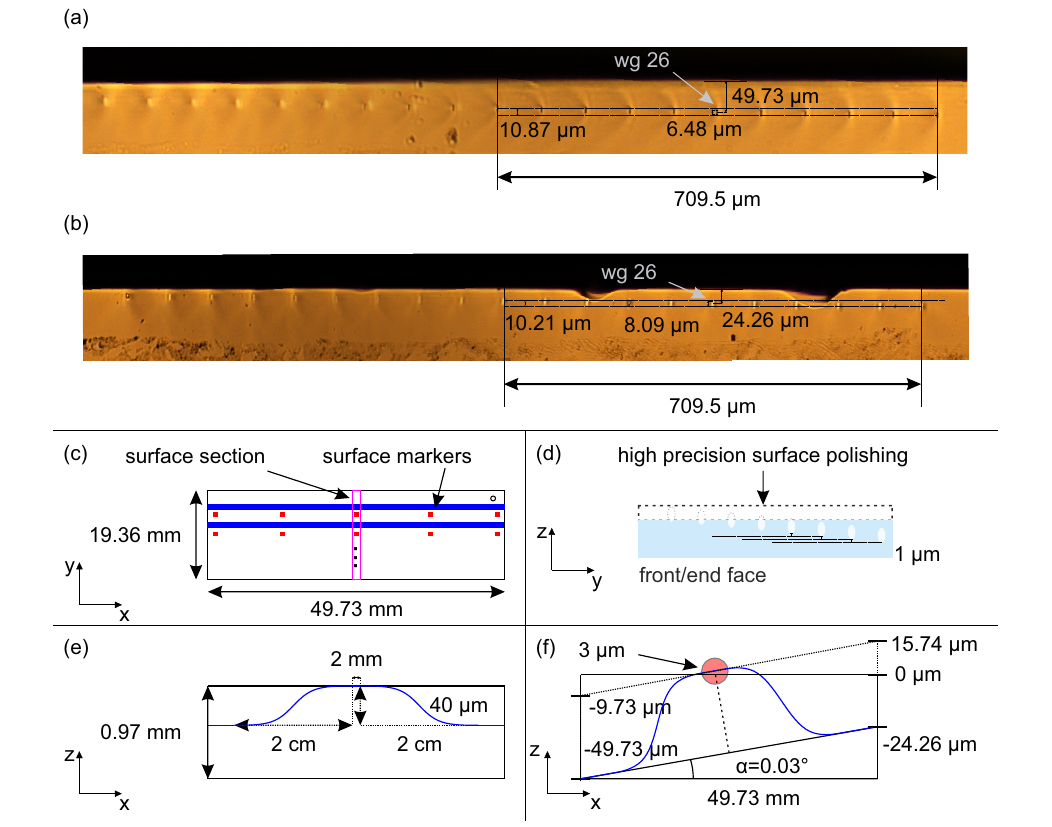}
	\caption{Dimensions of the waveguide chip and principle of the surface waveguides (a) Front face of the chip A with information on the distances between the waveguides, the distance to the upper edge, the height of the waveguides and their offset. (b) Rear face of chip A with equivalent information. (c) Glass chip dimensions and writing pattern of the waveguides in the chip with surface section in the middle and height markers, marked red here. (d) Schematic of front/end face with cascade arrangement of the waveguide structure. (e) Dimensions of the surface waveguide sections. (f) Calculated tilt with the determined values from (a) and (b).}
	\label{Supplement_fig_chip_principle}
\end{figure*}

\noindent
From Figure~\ref{Supplement_fig_chip_principle} it seems that waveguide 26 is located 3\,µm above the surface which would mean that the surface waveguide is mostly cut and there is no cladding. The same holds for waveguide 36 on chip 4. Apparently, even a waveguide that is cut in half still provides mode guiding. Determining the first completely polished waveguide with an optical microscope and counting the waveguides results in a waveguide thickness of $\sim$5\,µm. Therefore, we consider the mean value of d$_{wg}$=4\,µm in our calculations. 

\newpage
\begin{figure*}[h]
	\centering
	\includegraphics[width=1.0\textwidth]{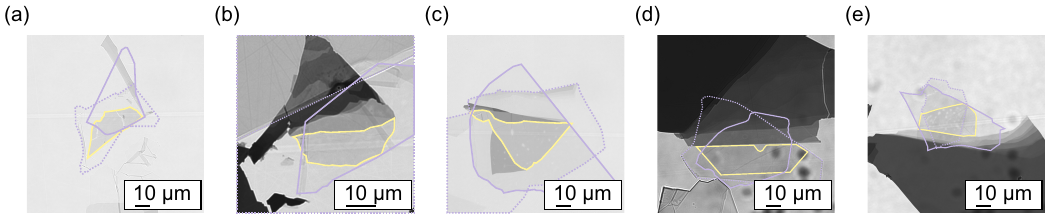}
	\caption{Images of the encapsulated monolayer samples on the waveguide. The TMD monolayer is framed in yellow and the hBN layers in purple. The top hBN layer is marked by dashed lines: (a) WSe$_2$ monolayer, sample length over waveguide $\sim$\,24\,µm, thin hBN encapsulation. (b) WS$_2$ monolayer, sample length over waveguide $\sim$\,28\,µm, thin hBN encapsulation. (c) MoSe$_2$ monolayer, sample length over waveguide $\sim$\,46\,µm, thin hBN encapsulation. (d) MoSe$_2$ monolayer, sample length over waveguide $\sim$\,48\,µm, thick hBN encapsulation. (e) Encapsulated WSe$_2$ monolayer, sample length over waveguide $\sim$\,23\,µm, thick hBN encapsulation.}
	\label{Supplement_fig_sample_pictures}
\end{figure*}

\noindent
Figure~\ref{Supplement_fig_sample_pictures} shows microscope images of the samples encapsulated in thin hBN in trans-illumination. The surface waveguide is visible underneath the samples.\\

\begin{figure} [h]
    \centering
    \includegraphics[width=1\linewidth]{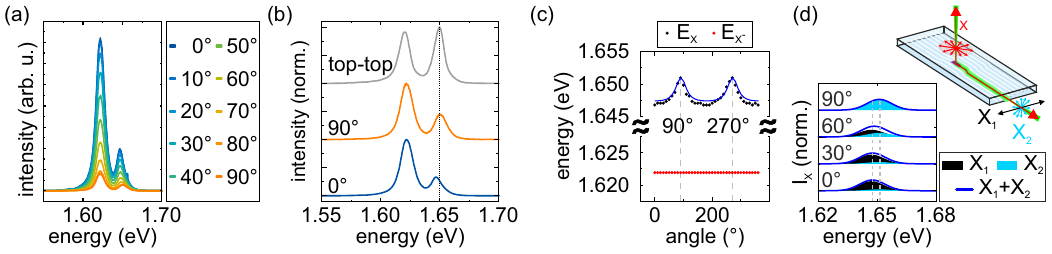}
    \caption{Angle-resolved low-temperature measurement results of MoSe$_2$ in top excitation and side collection measurement geometry. (a) µPL spectra with steps of 10$^\circ{}$ between in-plane  (0$^\circ{}$) and out-of-plane (90$^\circ{}$) detection polarization. (b) Comparison of spectral form of µPL spectra in top-top measurement geometry (grey) and top-side measurement geometry. (c) Angle dependence of the energy of trion and exciton peak. (d) Model concept: Emission of two gaussian components with different polarization dependence resulting in a shift of the exciton energy under different detection polarization angles. The blue line in c) represents the angle dependence of the A\:exciton peak energy calculated with this model concept.}
    \label{Supplement_fig_MoSe2_angle_series}
\end{figure}

\noindent
Figure~\ref{Supplement_fig_MoSe2_angle_series} shows results for MoSe$_2$ encapsulated in thin hBN. It highlights the angle dependence of the intensities of exciton and trion as well as of the energies. Regarding the intensities in Figure~\ref{Supplement_fig_MoSe2_angle_series}a without considering the energy shift we determine a degree of polarization of 0.77(1$\pm$5\%) for the exciton and 0.77(1$\pm$3\%) for the trion, which means that the signal is mainly, but not entirely linearly polarized.\\
Figure~\ref{Supplement_fig_MoSe2_angle_series}d shows that the polarization dependent shift of the A\,exciton energy is not a shift of the whole A\,exciton peak. The behavior of all measured parameters indicates that the peak for the A\,exciton must consist of two closely spaced components that exhibit different dependencies on polarization.
We suggest that X$_1$ represents the emission of an exciton coupled to the waveguide mode. According to our calculations exciton-polaritons form when combining an excitonic material and a waveguide. X$_2$ might be the emission of an uncoupled exciton that can emerge when the coherence length of the exciton is much smaller than the wavelength \cite{burstein_optical_1995}.

\newpage

\begin{figure} [h]
    \centering
    \includegraphics[width=1\linewidth]{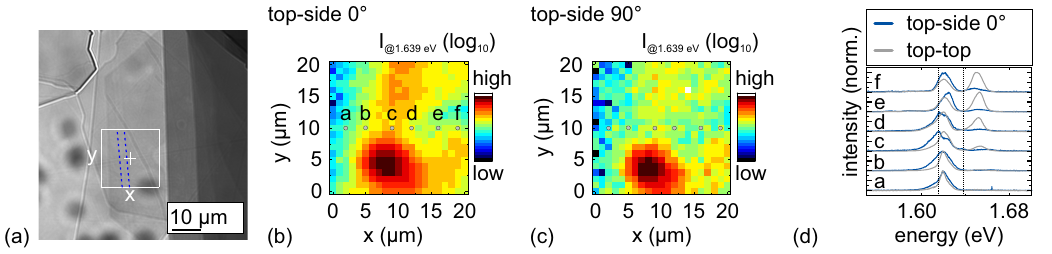}
    \caption{Low-temperature measurement results of MoSe$_2$ in top excitation and side collection measurement geometry: (a) Sample picture with indicated scan area (white) and waveguide (blue). (b) False color map of PL intensity at the energy position 1.639 eV for horizontal detection polarization. (c) False color map of PL intensity at the energy position 1.639 eV for vertical detection polarization. (d) Normalized PL spectra as indicated resulting in a spectral profile perpendicular to the waveguide, comparison with corresponding spectra in top-top measurement geometry.}
    \label{Supplement_Results_MoSe2_scans_thick_hBN}
\end{figure}

\begin{figure} [h]
    \centering
    \includegraphics[width=1\linewidth]{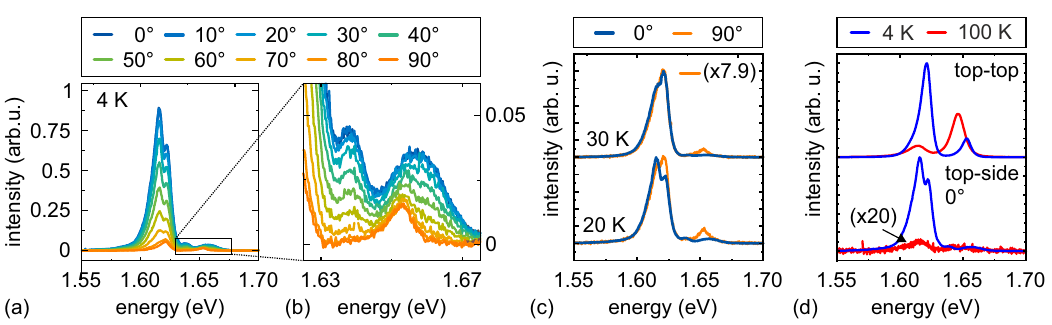}
    \caption{Angle- and temperature series measurement results of MoSe$_2$ encapsulated in thick hBN: (a,b) µPL spectra with steps of 10$^\circ{}$ between in-plane  (0$^\circ{}$) and out-of-plane (90$^\circ{}$) detection polarization with additional focus on the A\:exciton section.  (c) Normalized PL spectra at 20\,K and 30\,K.  (d) Comparison of normalized spectra at 4\,K and 100\,K in top-top and top-side measurement geometry.}
    \label{Supplement_Results_MoSe2_temperature_series}
\end{figure}

\begin{figure} [h]
    \centering
    \includegraphics[width=1\linewidth]{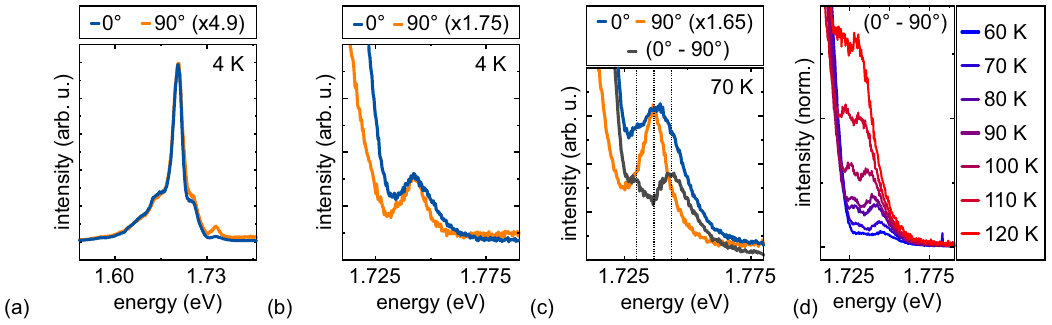}
    \caption{Low-temperature measurement results of WSe$_2$ encapsulated in thick hBN layers in top excitation and side collection measurement geometry. (a) Spectra in horizontal polarization (blue) and vertical polarization (orange) at a temperature of 4\,K. (b) The same spectrum as in a) but with focus on the exciton spectral range. (c) Spectra at 70 K with focus on the exciton spectral range. (d) Temperature series of in-plane spectra with subtracted out-of-plane signal focusing on the A exciton spectral range.}
    \label{Supplement_Results_WSe2_thick_hBN}
\end{figure}

\noindent
For WSe$_2$ monolayers encapsulated in $\sim$40\,nm thick hBN layers, displayed in Figure~\ref{Supplement_Results_WSe2_thick_hBN}a, a suppression of the exciton intensity is noticeable as it was for MoSe$_2$ in the main text. A splitting as well-defined as for MoSe$_2$ could not be observed, probably because of several other lower energy peaks like the trion and defect exciton peaks that overlay the lower polariton peak. A lower energetic feature with an energy difference of 13.9\,meV as observed for MoSe$_2$ would already be covered by the trion. Furthermore, based on the measurements with thinner hBN encapsulation in Figure~\ref{Panel2_TMD_spectra} of the main text and the larger oscillator strength of WSe$_2$ \cite{leppenen_exciton_2020} an even larger splitting would be expected, which would be even more difficult to detect.\\
We observe a slight broadening of the exciton peak towards higher energies in horizontal detection polarization, displayed in \ref{Supplement_Results_WSe2_thick_hBN}b. In analogy to the measurements for MoSe$_2$, this could indicate an upper polariton feature.\\ 
At a temperature of 70\,K, in Figure~\ref{Supplement_Results_WSe2_thick_hBN}c, a slight shoulder emerges which could be attributed to a lower polariton with a splitting of 6.9\,meV to the uncoupled exciton. Similar to the MoSe$_2$ measurements the out-of-plane detected spectrum can be used as background which provides a clearer image of the splitting. Up to 120\,K the splitting described in this way remains visible, but is reduced with temperature. This is comparable to the results by Zhang et al. who determined the transition to the weak coupling regime at around 130\,K for WSe$_2$ on a photonic crystal waveguide \cite{zhang_photonic-crystal_2018}.

\newpage
\section{Supplementary theoretical data}

\begin{figure*}[h]
	\centering
	\includegraphics[width=1.0\textwidth]{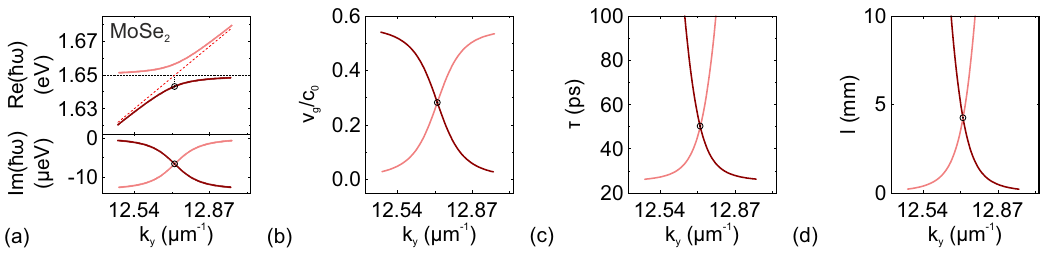}
	\caption{Transfer matrix calculations for MoSe$_2$ with 39.93 nm (121 layers) hBN thickness of top and bottom hBN. dark red: lower polariton mode, light red: upper polariton mode, bottleneck indicated with a circled black diamond. (a) Top: real part of the dispersion relation as in Figure~\ref{Panel3_Dispersion_Calculations}a in the main text, dashed bright red line: stacked multilayer without TMD. Bottom: Imaginary part of the dispersion relation. (b) Group velocities of the lower- and upper polariton branch. (c) Polariton lifetimes of the lower- and upper polariton branch.(d) Polariton propagation lengths of the lower- and upper polariton branch.}
	\label{Supplement_fig_theory_group_velocity}
\end{figure*}

\noindent
The group velocity in Figure~\ref{Supplement_fig_theory_group_velocity} was calculated with

\begin{equation*}
    v_g=\frac{d Re(\omega)}{dk_y}
\end{equation*}
and the polariton lifetime with
\begin{equation*}
    \tau=\frac{1}{|2Im(\omega)|}.
\end{equation*}
By combining these the ideal polariton propagation length can be calculated \cite{kavokin_cavity_2003,beggs_waveguide_2005}:
\begin{equation*}
    l=v_g\tau.
\end{equation*}

\begin{figure*}[t!]
	\centering
	\includegraphics[width=1.0\textwidth]{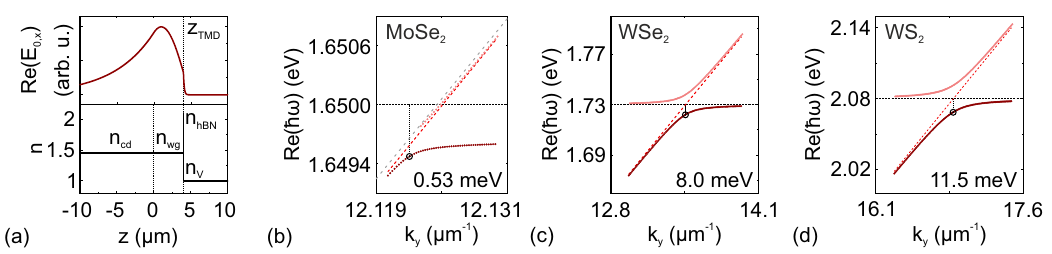}
	\caption{Transfer matrix calculations for MoSe$_2$, WSe$_2$ and WS$_2$: (a) Bottom: refractive index profile of the structure with 8.25\,nm (25 layers) hBN thickness of each top and bottom hBN, top: electric field profile of the TE-mode of the lower polariton branch at the bottleneck for MoSe$_2$. (b) Dispersion relation of the structure in a): Bright red line: stacked multilayer without TMD, dark red: lower polariton mode, bottleneck indicated with a black diamond, light red: upper polariton mode, dashed grey line: light line of $k_y=n_{cd}\frac{\omega}{c}$, the condition for total internal reflection and formation of a bound mode. (c) Dispersion relation of WSe$_2$ with encapsulation in 39.93\,nm hBN top and bottom layers and indicated splitting. (d) Dispersion relation of WS$_2$ with encapsulation in 39.93\,nm hBN top and bottom layers and indicated splitting.}
	\label{Supplement_fig_theory_dispersion}
\end{figure*}

\begin{figure*}[t!]
	\centering
	\includegraphics[width=1.0\textwidth]{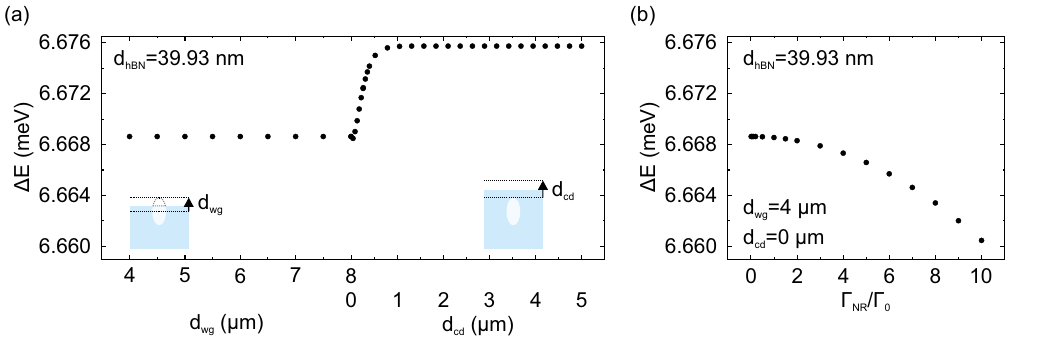}
	\caption{Dependence of the energy difference of the lower polariton to the exciton resonance energy on different parameters of the calculation: (a) Dependence on the waveguide thickness for a cut waveguide and dependence on the cladding thickness for an embedded waveguide (b) Dependence on the ratio of non-radiative broadening to radiative broadening}
	\label{Supplement_fig_theory_thickness}
\end{figure*}

\newpage
\subsection{Discussion of the applied calculation parameters}
\label{subsec_explanations_parameters}

\noindent
$n_{\text{hBN}}$ was calculated with the respective $E_A$ for each material as \cite{lee_refractive_2019}:
\begin{equation*}
    n=\sqrt{\frac{3.263 \lambda^2}{\lambda^2- \left(0.1644 \cdot 10^{-6} \text{m} \right)^2} +1}
\end{equation*}
$d_\text{wg}$ was estimated by analyzing the height markers and the optical microscope pictures of the end facets and the top view (see Figure \ref{Supplement_fig_chip_principle}). The full width of the waveguide would be $\sim$8\,µm, but the analysis showed that the used waveguide is polished to around half of its width, yet still guides light.\\
$d_\text{hBN}$ is an estimation regarding the reflection on PDMS.
$E_A$ is the A\,exciton energy of the considered TMD monolayer.
$\tau_0^{LT}$ is the radiative lifetime of the exciton at room temperature \cite{Palummo_exciton_2015}. Is is used to calculate the required radiative broadening $\Gamma_0$ calculated as \cite{kavokin_cavity_2003}:
\begin{equation*}
    \Gamma_0=\frac{1}{2\tau_0}
    \label{eq_radiative_broadening}
\end{equation*}
The nonradiative broadening is estimated. Following Gupta et al. it needs to be smaller than the radiative broadening at low temperatures \cite{gupta_fundamental_2019}.

\end{document}